\documentclass{iaus}
\usepackage{graphicx}

\title[TMT and Lensing]{Thirty Meter Telescopes and Gravitational Lensing}

\author[R. G. Carlberg]
{R. G. Carlberg
}

\affiliation{
Department of Astronomy and Astrophysics, University of Toronto,
Toronto, Ontario M5S 3H8, Canada email: carlberg@astro.utoronto.ca}

\pubyear{2004}
\volume{225}
\pagerange{1--8}
\date{?? and in revised form ??}
\jname{Impact of Gravitational Lensing on Cosmology}
\editors{Mellier, Y. \& Meylan,G. eds.}
\begin{document}

\maketitle

\begin{abstract}
Diffraction limited 30m class telescopes will play an important role
in gravitational lensing studies, coming online in approximately 2015.  As
imaging telescopes they will complement the $\sim$6m JWST, probing to
smaller angular scales in greatly magnified objects near critical
lines and for measuring shear of objects below the JWST angular scale,
such as luminous super-star clusters at high redshift. The high source
density will allow more detailed mass mapping in the weak lensing
regime and will be useful in breaking the cosmology-lens potential
degeneracy in strong lensing.  As multi-object spectrographs 30m
telescopes should provide spectra over the entire optical and near
infrared spectrum region. The statistical distribution of redshifts
needed to invert projected shear measurements and calibration of
photometric redshifts for ``tomography'' will be available to flux
levels around 5-10 nano-Jansky (approx 29.5 m$_{AB}$). However, a one nJy
object is expected to require $\sim$500 hours to acquire a redshift,
which is most of the dark time in an observing season. Accordingly
``gravitational telescopes'' will be an important tool for probing the
very faint high redshift universe, magnifying a few square arc-seconds
at a time by factors of 10-1000.
\end{abstract}

\firstsection 

\section{Introduction}

The science cases for 30 meter class telescopes emphasize that there
are important problems of faint objects, small galaxies at high
redshift, star formation and planet formation where we need to have
diffraction limited imaging and spectrographs and increased
spectrographic capabilities in the near-infrared, without giving up
the undoubted benefits of the optical part of the spectrum. The new
capabilities will have important benefits for the study of
gravitational lensing to probe problems in cosmology, galaxy formation
and the distribution of lensing mass. In particular, the capabilities
of examining smaller, fainter source populations will allow a much
higher sky density of background sources to be used, allowing more
precise mass mapping and providing sufficient redundancy that the
mass-cosmology degeneracy can be resolved.

The performance of a 30m telescope will be considered for its utility
with lensing problems. In the case of weak lensing, images useful for
shear analysis and photometric redshifts are already available from
HST well beyond the limit of calibrated (photometric) redshift
distributions. The image/spectra mismatch will widen with the arrival
of JWST if only 10m class telescopes are available. The $D^4$
advantage of a diffraction limited telescope provides nearly two
orders of magnitude gain at 30m. For strong lensing, the imaging
capabilities of a 30m telescope will allow the smallest angular scales
to be resolved which will preserve regions of very high
magnification. To obtain even an R=2000 spectrum for sources fainter than
about five nJy will likely be routine through the study of sources
that are magnified by a factor of ten or more in a strong lens. Of
course this approach requires that the source population be
sufficienty numerous, more than about $10^5$ per square degree, that
there will be substantial numbers. 

Adaptive optics is not without its limitations. The PSF has a
diffraction limited core plus a ``halo'' with the characteristics of
the natural seeing PSF. To a first approximation the light in the
diffraction core is given by $S$, the Strehl ratio. For isolated
sources this means that performance scales as $S^2 D^4$, where $D$ is
the aperture of the telescope. However, for high surface brightness
backgrounds and crowded regions the performance can scale as $S^4
D^4$, placing a high premium on a high quality Strehl in the AO
system.

\begin{figure}
        \includegraphics[width=\hsize]{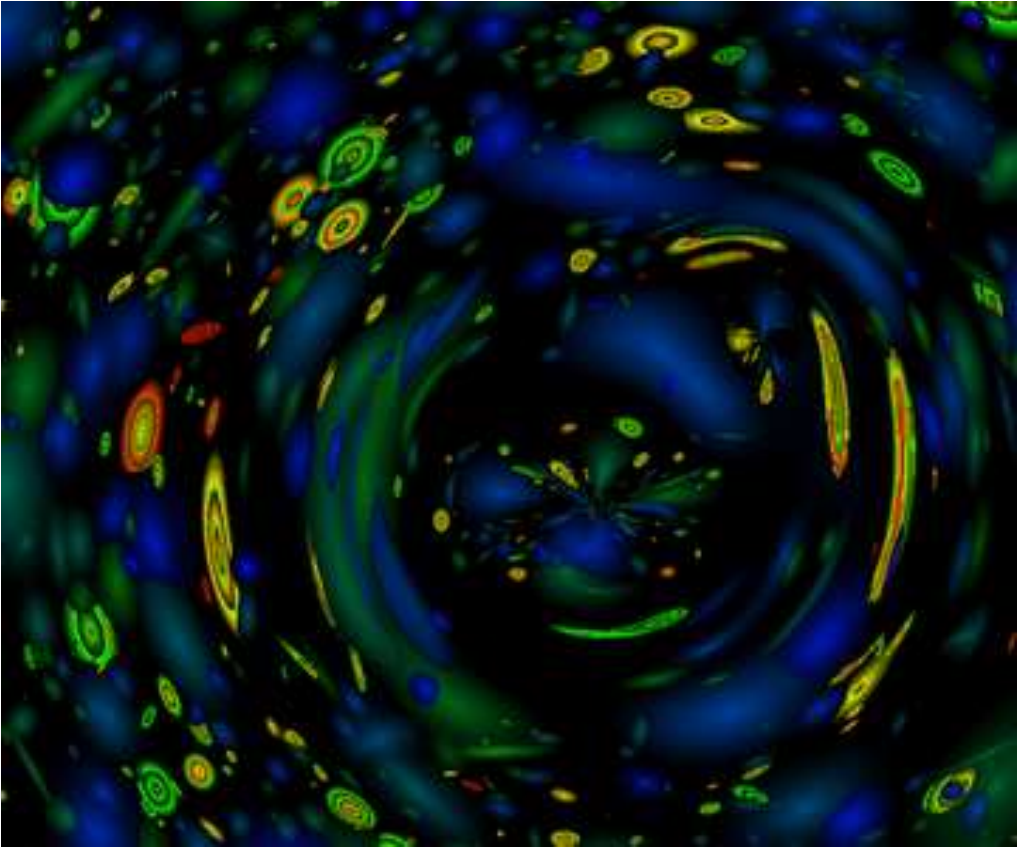}
\caption{A simulation of a $40\times32^{\prime\prime}$ piece of sky with galaxies visible to redshift
fifteen (the skymaker simulator from Terapix with a speculative
luminosity and mass evolution!) visible through a 800 km~s$^{-1}$
elliptical lens with an offset 300 km~s$^{-1}$ lens. The colors
reflect the redshift, blue at zero, yellow at $z=6$ and red at
$z=15$. The intensity scale is logarithmic. In a region of say 10x10
arcsec, there are dozens of objects, spanning a wide range of redshifts
hence have significantly different shear and magnification near the same image plane
locatoin.}
\end{figure}

\section{Weak lensing analysis and 30m telescopes}

Hubble Space Telescope imaging in the Ultra Deep Field can reach to
$m_{AB}=$ 28.5 to 29.0 mag \cite{Beckwith03}. The $\sim6$m aperture of
JWST will make these depths routine, particularly at wavelengths
beyond 1 micron, which will allow much larger sky areas to be acquired
(although still at considerable expense) as is required for many
aspects of lensing analysis. Although photometric redshifts are a
powerful tool to determine redshifts to the precision required for
statistical inversion of lensing measurements (such as cosmic shear or
higher moment measurements) they are secured on the basis of a
spectroscopic calibration of the redshifts. The ``Generic GSMT''
exposure time calculator (http://www.noao.edu/noao/staff/brooke/gsmt/)
finds that it takes about $7.8\times 10^5$ seconds (about 43 dark
nights of 5 hours) to obtain a $s/n=5$ {\it continuum} spectrum at
$1.6\mu$ with a 30m telescope and an R=2000 spectrograph operating
near the diffraction limit ($0.03\times 0.15^{\prime\prime}$ slit)
with a Strehl-throughput product of 0.3. Although this is a very
expensive observation, it is comparably costly in the limited-lifetime
JWST mission time (see exposure times at
http://www.stsci.edu/jwst/science/jms/reports.html). The TMT project
plans to be able to acquire spectra to this depth over a field of
about 5 arcminutes, which will in principle allow hundreds of spectra
to be obtained at the same time.

Thirty meter class telescopes are unlikely to be very useful for any
wide area imaging work, although it is not specifically
precluded. Large areas of natural seeing imaging are best done on
smaller aperture telescopes. Space-based telescopes will provide
enviable PSF stability over relatively large fields. However, for
targeted observations the TMT will have high quality AO imaging
capabilities at the diffraction limit over a field of about one
arcminute and sampling of of a few milli-arcseconds. This will be of
interest for targeted studies on the scales of individual galaxy
halos, for which 100 kpc subtends about a 20 arcsecond circle at
redshift one. At a source density of $10^6$ per square degree there
will be approximately one hundred field galaxies per halo, of which at
least half will be background for lens galaxies at redshift one. There
is also the prospect that components of background galaxies, such as
the nuclear bulge and compact HII regions will make suitable sources
for statistical shear measurements. In principle these measurements
could be of interest for galaxy size halos over approximately redshift
0.5 to 5, becoming too large at low redshift and too few bright
background objects at high redshift.

\section{Strong Lensing}

Strong gravitational lensing studies will be revolutionized on 30m
class telescopes. The detection of lensing events is strongly
dependent on the telescope aperture and angular resolution.  First,
the probability of lensing increases in direct proportion to the
density of background galaxies that a telescope can comfortably
reach. Second, the telescope must be able to resolve the arc,
otherwise it dilutes the high surface brightness of very small or
unresolved regions, such as super-star clusters and nuclear bulges at
high redshift. Four meter class telescopes at good sites (and in
space!) opened up gravitational lensing because they were the first to
have sub-arcsecond images and the aperture to comfortable reach
sources densities of about $10^4$ per square degree. At sky densities
approaching $10^6$ degree$^{-2}$ the faint background galaxies are
near (or below) the angular resolution of the Hubble Space telescope
in the optical bands. Such images are well matched to the angular
resolution in the K band available from JWST and AO-equipped
ground-based telescopes.

\begin{figure}
        \includegraphics[width=0.5\hsize]{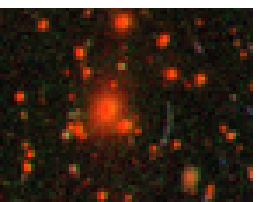}
        \includegraphics[width=0.48\hsize]{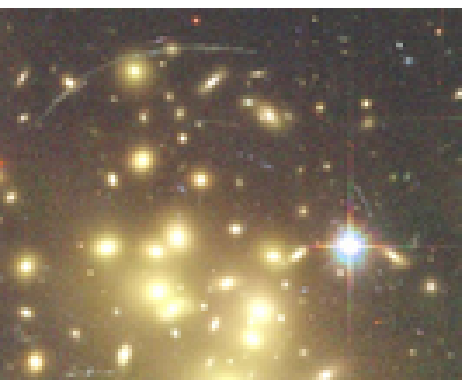}
\caption{Left: a discovery image of a gravitational arc from the CFHT Legacy survey. 
Right: Part of the spectacular Hubble ACS image of Abell 1689,
emphasizing the importance of having near diffraction limited
performance to avoid diluting unresolved high surface brightness
features. The image size ratio between these two images, about a
factor of 7, is roughly the improvement that a 30m will give in the
near-IR, with spectroscopy likely being one of the main uses. The
bright star in the lower right of the HST image is the guide star at
the center of Figure 3.}
\end{figure}
 
A ground-based 30m class telescope with AO should reveal many hundreds
and possibly thousands of arcs in galaxy clusters (see Fig 1), and
should make multiple images a much more common phenomena for
individual galaxies.  In galaxy clusters an important consequence is
that within a region of say $10x10$ arcseconds there will be several
background galaxies with varying shear, depending on their redshift
and precise position relative to the caustic surfaces. This opens up
the opportunity of locally breaking the degeneracy between the
gradient of the lensing potential and the cosmology if redshifts are
available for all the sources as well as the lens. In the case of
smooth lenses with simple elliptical asymmetry this would be a simple
and powerful tool. However, in the presence of irregularities in the
cluster potential, such as the visible galaxies and remnant (dark
matter) tidal tails, statistical averaging will be required to derive
high precision cosmological information. However, since the
measurements are primarily geometrical, position and shape of very
high precision and the ``noise'' is of astrophysical interest as well,
the 30m telescopes will open a new domain of investigation. Moreover,
since the probability of lensing is linear in the density of
background sources, every galaxy cluster will exhibit at least a few
arcs. For a 10:1 arc the source plane area magnified is about one
tenth the radius of the critical line in the cluster, typically 15
arcsec. Therefore the magnified area is about 5 sq arcseconds. But,
with several thousand massive clusters, the total area magnified
becomes large enough to create a very useful sample.

\begin{figure}
        \includegraphics[width=0.45\hsize]{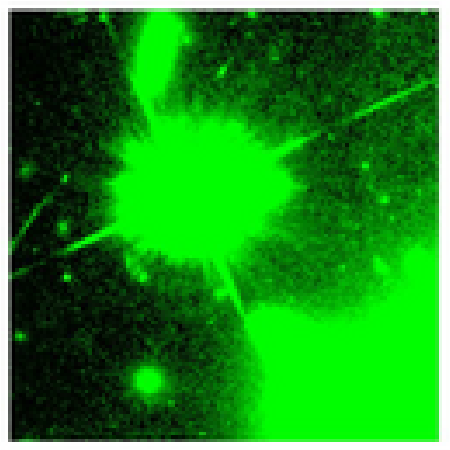}
~~~~~~~~~~~
        \includegraphics[width=0.45\hsize]{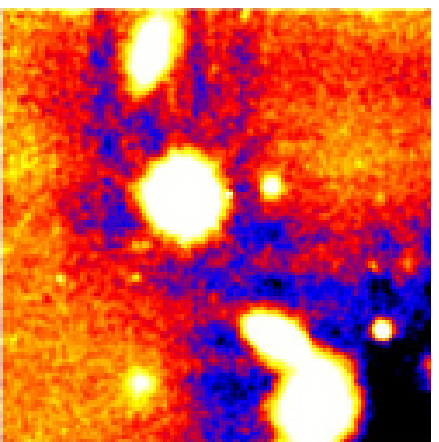}
\caption{The region of the bright star ($m_V\simeq 13$ mag) in A1689. Left: the 
ACS image i band image. The field is $16^{\prime\prime}$ on a
side. Precisely the same area done with the Gemini Altair adaptive
optics system in the K band in a 3 hour image. The limiting magnitude
is $K_{AB}\simeq 26$. The PSF core is about 0.08 arcseconds in both
images. The faint arc to the left of the guide star is detected in
K. Most of the faint galaxies are also visible in K light. MCAO
systems on 8m telescopes will enlarge the field of view and provide a
more uniform PSF. However, these images already show the limitations
of 8m class telescope AO for strong lensing studies. The $D^4$ gain,
along with increased field, on an 30m will be transformative.}
\end{figure}

\section{Gravitational Telescopes}

It is clear that routinely reaching to nano-Jansky flux levels for
high redshift galaxies and ``first light'' objects is just within the
reach of of JWST and TMT. A single very massive ``first star'' is
likely to appear at about $m_{AB}\simeq 35$ or so at redshift 15,
however a super star cluster may consist of some dozen or so of these
stars and much of the light may emerge from an HII region near the
resolution limit of a 30m. Young star clusters in galaxies like the
Antennae (see Figure 4) that are near cluster critical lines will be
massively amplified. Given that 2-8m routinely find factors of ten,
the superior PSF of a 30m should frequently (the numbers are roughly
inversely proportional to the amplification) find magnifications of a
factor of a hundred. The resulting five magnitude brightness boost, as
well as the one-dimensional increase in linear scale will be important
to allow the study of a representative sample of very faint, very high
redshift sources. Almost all of this work will be in the JHK bands,
where one can work between the emission lines of the sky spectrum to
take advantage of the relatively dark sky in between.

\begin{figure}
        \includegraphics[width=\hsize]{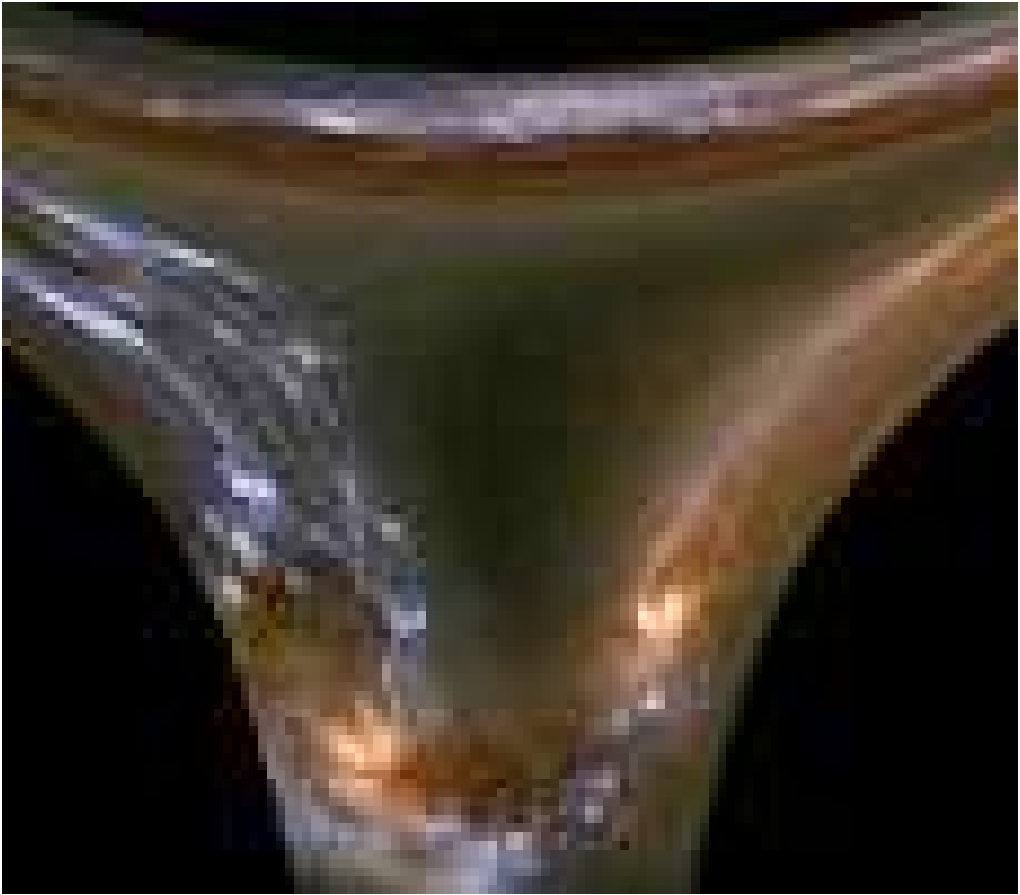}
\caption{The Antennae galaxies scaled to redshift three 
and placed across the critical line of a 800 km~s${-1}$ galaxy
cluster. The field of view is 3 arcseconds. The super-star clusters
are not resolved in this image, The SSCs near the critical line
will be magnified by factors around one hundred. }
\end{figure}

\section{Project status}

The combination of scientific importance and technical excitement has
lead to several projects to advance 20-30m class projects, and
larger. One example is the Thirty Meter Telescope Project (TMT), a
partnership of Caltech, the University of California, AURA and ACURA
(Canada). At the time of writing the TMT project plans to go to cost
review (i.e. a technically and financially complete project) in
mid-2006. There is about 40 FTE of activity in TMT at present which
will rapidly rise over the next few months in order to meet this
milestone and continue on schedule to construction.

\section{Conclusions}

Thirty meter class telescopes will be important partners for
gravitational lensing studies. They will primarily provide the
spectroscopic redshifts to calibrate photometric redshifts of deep
imaging surveys. A 30m will also be useful for deep imaging studies on
the scales of individual galaxy halos, over the redshift of 0.5 to 5
range. Strong lensing will be one of the few ways to obtain spectra of
objects below a few nano-Jansky (unlensed) level at levels where the
spectra can be adequately dispersed for astrophysical
study. Diffraction limited spectroscopy using a 30m is particularly
well suited to this exciting area of study.

\begin{acknowledgments}
Support for the Canadian Large Optical Telescope work comes from
NSERC, NRC and the Canada Foundation for Innovation. The A1689 AO
image is a collaboration amongst Tim Davidge, Stephen Gwyn, Luc Simard
and RGC using Gemini.  Thanks to members of the Canadian LOT SAC, the
TMT SAC, the GSMT Science Working Group, and the European ELT Science
Working Group for lively ongoing discussions.
\end{acknowledgments}

\begin{discussion}

\discuss{Danielle Alloin}
{What is the cost estimate for instrumentation for a 30m telescope?}

\discuss{Carlberg}{The currently available cost studies
were done at the conceptual design stage. The total cost of a 30m is
estimated as \$700-800M US dollars, which includes a contingency of
nearly \$100M. The estimated ``first light'' instrumentation budget is
\$80M, with a separate budget of approximately \$150M for adaptive
optics. It is currently planned that there will be three first light
instruments built within this budget.  A PDR level cost review should
be available in mid-2006 from TMT.}

\end{discussion}

\end{document}